\documentclass[preprint,aps,prc,tightenlines,floatfix,showpacs,showkeys,nofootinbib]{revtex4}

\usepackage{graphicx}
\usepackage{dcolumn}
\usepackage{bm}
\usepackage{longtable}

\newcommand{\be}{\begin{equation}}
\newcommand{\ee}{\end{equation}}
\newcommand{\bea}{\begin{eqnarray}}
\newcommand{\eea}{\end{eqnarray}}

\begin{document}
\title{Neutrinoless double beta decay of deformed nuclei within QRPA with realistic interaction}

\author{Dong-Liang Fang}
\affiliation{Institut f\"ur Theoretische Physik, Universit\"at T\"ubingen, 
D-72076 T\"ubingen, Germany}
\author{Amand Faessler}
\affiliation{Institut f\"ur Theoretische Physik, Universit\"at T\"ubingen, 
D-72076 T\"ubingen, Germany}
\author{Vadim Rodin}
\email{vadim.rodin@uni-tuebingen.de}
\affiliation{Institut f\"ur Theoretische Physik, Universit\"at T\"ubingen, 
D-72076 T\"ubingen, Germany}
\author{Fedor \v Simkovic}
\affiliation{BLTP, JINR, Dubna, Russia and Department of Nuclear Physics, Comenius University, SK-842 15 Bratislava, Slovakia}

\date{\today}

\begin{abstract}
In this paper a microscopic state-of-the-art approach 
to calculation of the nuclear matrix element $M^{0\nu}$ for neutrinoless double beta decay  with an account for nuclear deformation is presented in length and applied for $^{76}$Ge, $^{150}$Nd and $^{160}$Gd.
The proton-neutron quasiparticle random phase approximation (QRPA) with a realistic residual interaction [the Brueckner $G$ matrix derived from the charge-depending Bonn (Bonn-CD) nucleon-nucleon potential] is used as the underlying nuclear structure model.  The effects of the short range correlations and the quenching of the axial vector coupling constant $g_A$ are analyzed. The results show that
neutrinoless double beta decay of $^{150}$Nd, to be measured soon by the SNO+ collaboration, provides one of the best probes of the Majorana neutrino mass.
This confirms our preliminary conclusion in Ref.~\cite{Fang10}.

\end{abstract}

\pacs{
23.40.-s, 
23.40.Bw 
23.40.Hc, 
21.60.-n, 
}

\keywords{Majorana neutrino; Neutrino mass; Double beta decay; Nuclear matrix element; Nuclear deformation}

\date{\today}
\maketitle

\section{Introduction}
Neutrinoless double beta decay ($0\nu\beta\beta$-decay) is a
second order nuclear weak decay process with the emission of two electrons
only~\cite{Vog92,fae98,AEE07}: 
$(A,Z)\rightarrow(A,Z+2)+2e^-$. 
This process is forbidden in the standard model (SM) of electroweak interaction since it violates the conservation of the total lepton number. 
The observation of $0\nu\beta\beta$ decay will immediately prove the neutrino to be identical to its antiparticle (a Majorana particle) as assured by the Schechter-Valle theorem~\cite{SV82}. Thereby, $0\nu\beta\beta$ decay offers the only feasible way to test the charge-conjugation property of the neutrinos. In addition, the existence of $0\nu\beta\beta$ decay requires that the neutrino is a massive particle.

The fact that the neutrinos have non-vanishing masses has firmly been established by neutrino oscillation experiments
(see, e.g., Ref.~\cite{Kay08}).
However, the oscillation experiments cannot in principle measure the absolute scale of the neutrino masses. One of the possible ways to probe the absolute neutrino masses at the level of tens of meV is to study $0\nu\beta\beta$ decay.
Provided the corresponding $0\nu\beta\beta$-decay rates are accurately measured, 
reliable nuclear matrix elements (NME) $M^{0\nu}$ will be needed
to  deduce the effective Majorana neutrino mass from the experimental half-lives of the decay. 

One of the best candidate for searching $0\nu\beta\beta$ decay is $^{150}$Nd since 
it has the second highest endpoint, $Q_{\beta\beta}=$3.37 MeV, and the largest phase space factor for the decay (about 33 times larger than that for $^{76}$Ge, see e.g.~\cite{Vog92}). 
The SNO+ experiment at the Sudbury Neutrino Observatory will use Nd loaded scintillator to search for neutrinoless double beta decay by looking for a distortion in the energy spectrum of decays at the endpoint~\cite{SNO+}. 
SNO+ will be filled with 780 tonnes of liquid scintillator. The planned loading of 0.1\%  of the natural Nd translates into 43.6 kg of the isotope $^{150}$Nd. It is expected to achieve the sensitivity of $T^{0\nu}_{1/2} \simeq 5\cdot 10^{24}$ years after one year of running, with the best final value of about 3--4 times longer (without enrichment of the dissolved Nd). With the NME $M^{0\nu}=4.74$ of Ref.~\cite{Rod05}, obtained within the proton-neutron quasiparticle random phase approximation (QRPA) with neglect of deformation, already the initial phase of SNO+ will be able to probe $m_{\beta\beta}\approx$ 100 meV, and will finally be able to achieve sensitivity of $m_{\beta\beta}\approx$ 50 meV corresponding to the inverse hierarchy (IH) of the neutrino mass spectrum.  

However, $^{150}$Nd is well-known to be a rather strongly deformed nucleus. This 
strongly hinders a reliable theoretical evaluation 
of the corresponding $0\nu\beta\beta$-decay NME
(for instance, it does not seem feasible in the near future to reliably treat this nucleus within the large-scale nuclear shell model (LSSM), see, e.g., Ref.~\cite{men09}). 
Recently, more phenomenological approaches like the pseudo-SU(3) model~\cite{Hir95}, the Projected Hartree-Fock-Bogoliubov (PHFB) approach~\cite{Hir08} and the interacting boson model (IBM-2)~\cite{Bar09} have been employed to calculate 
$M^{0\nu}$ for strongly deformed heavy nuclei (a comparative analysis of different approximations involved in the models can be found in Ref.~\cite{Esc10}).
The results of these models generally reveal a substantial suppression of $M^{0\nu}$ for $^{150}$Nd as compared with the QRPA result of  Ref.~\cite{Rod05} where $^{150}$Nd and $^{150}$Sm were treated as spherical nuclei. However, the calculated NME $M^{0\nu}$ for $^{150}$Nd reveal rather significant spread.

The most microscopic way up-to-date to describe the effect of nuclear deformation on 
$\beta\beta$-decay NME $M^{2\nu}$ and $M^{0\nu}$ is provided by the QRPA. In Refs.~\cite{Sim03,Sim04,Sal09} a QRPA approach for calculating 
$M^{2\nu}$ in deformed nuclei has been developed.  Theoretical interpretation of
the experimental NME $M^{2\nu}_{exp}$ which have been obtained for a dozen of nuclei~\cite{Bar10} provides a test of different theoretical methods. 
It was demonstrated in Refs.~\cite{Sim03,Sim04,Sal09} that 
deformation introduces a mechanism of suppression of the $M^{2\nu}$ matrix element 
which gets stronger when deformations of the initial and final nuclei differ from each other. A similar dependence of the suppression of both $M^{2\nu}$ and $M^{0\nu}$ matrix elements on the difference in deformations has been found in the PHFB~\cite{Hir08} and the LSSM~\cite{men09}. 

In the previous rapid communication~\cite{Fang10} we reported on the first QRPA
calculation of $M^{0\nu}$ for $^{150}$Nd with an account for nuclear deformation. 
The calculation showed a suppression of $M^{0\nu}$ by 
about 40\% as compared with our previous QRPA result for $^{150}$Nd~\cite{Rod05} that was obtained with neglect of deformation.
In this paper we give the details of the calculation of Ref.~\cite{Fang10}, and also include different nuclei in the analysis. In addition, the effects of the short range correlations and the quenching of the axial vector coupling constant $g_A$ are considered.
Making use of the newest NME, one may conclude that $0\nu\beta\beta$ decay
of $^{150}$Nd, to be searched for by the SNO+ collaboration soon, provides one of the best sensitivities to the Majorana neutrino mass and may approach the IH region of the neutrino mass spectrum.

\section{Formalism}

For the light Majorana neutrino exchange mechanism,
the inverse $0\nu\beta\beta$-decay lifetime is given by the product of three factors,
\begin{equation}
\label{3fact}
\left(T^{0\nu}_{1/2}\right)^{-1}=G^{0\nu}\,\left|{M'}^{0\nu}\right|^2\,m_{\beta\beta}^2\ ,
\end{equation}
where $G^{0\nu}$ is a calculable phase space factor, ${M'}^{0\nu}$ is 
the $0\nu\beta\beta$ nuclear matrix element, and $m_{\beta\beta}$ is the (nucleus-independent) ``effective Majorana neutrino mass'' which, in standard notation \cite{PDGR}, reads
\begin{equation}
m_{\beta\beta}=\left|\sum_{1=1}^3 m_i\,U^2_{ei}\right|\ ,
\end{equation}
$m_i$ and $U_{ei}$ being the neutrino masses and the $\nu_e$ mixing matrix elements,
respectively.
The NME includes both Fermi (F) and Gamow-Teller (GT) transitions, plus a small tensor (T) contribution~\cite{fae98},
\begin{equation}
\label{M'}
{M'}^{0\nu} = \left(\frac{g_A}{1.25}\right)^2
\left(-M^{0\nu}_F\frac{g^2_V}{g^2_A} + M^{0\nu}_{GT} +  M^{0\nu}_T
\right)\ .
\end{equation}
In the above expression, $g_A$ is the effective axial coupling in nuclear matter, not necessarily equal to its ``bare'' free-nucleon value $g_A\simeq1.25$. We note that for $g_A=1.25$  nuclear matrix element ${M'}^{0\nu}$ coincides with the standard definition ${M}^{0\nu}=-M^{0\nu}_F\frac{g^2_V}{g^2_A} + M^{0\nu}_{GT} +  M^{0\nu}_T$.
With the conventional prefactor $\propto g_A^2$ in Eq.~(\ref{M'}), the phase space 
$G^{0\nu}$ becomes independent of $g_A$.

The NME $M^{0\nu}$ for strongly deformed, axially-symmetric, nuclei can be most conveniently calculated within the QRPA in the intrinsic coordinate system associated with the rotating nucleus. This employs the adiabatic Bohr-Mottelson approximation that is well justified for $^{150}$Nd, which indeed reveals strong deformation. As for $^{150}$Sm, the enhanced quadrupole moment of this nucleus is an indication for its static deformation. Nevertheless, the experimental level scheme of $^{150}$Sm does not reveal a clear ground-state rotational band. In this work we treat $^{150}$Sm in the same manner as $^{150}$Nd. 
However, a more elaborated theoretical treatment going beyond the simple adiabatic approximation might be needed in the future to describe the nuclear dynamics of this nucleus.

The adiabatic, or the strong coupling (see, e.g., the monograph~\cite{RingSchuck80}), approach fails, however, for small deformations since the Coriolis force gets large and mixes states with different $K$. 
Nuclei $^{76}$Ge and $^{76}$Se have rather small deformations and the so-called weak coupling, or no alignment, limit~\cite{RingSchuck80} seems to be more suitable. In this limit the Coriolis force becomes so strong that the angular momenta of the valence nucleons get completely decoupled from the orientation of the core. 
Such a case would deserve a separate detailed study,
and the adiabatic approach to description of excited states of deformed nuclei is adopted in the present application of the QRPA with a realistic residual interaction. Nevertheless, one might already anticipate without calculations that in the weak coupling limit the calculated observables should reveal smaller deviations from the ones obtained in the spherical limit than those calculated in the strong coupling limit of the present work. In this connection it is worth to note that spherical QRPA results can exactly be reproduced in the present calculation by letting deformation vanish, in spite of the formal inapplicability of the strong-coupling ansatz for the wave function in this limit.

Nuclear excitations in the intrinsic system $| K^\pi\rangle$ are characterized by the projection of the total angular momentum onto the nuclear symmetry axis $K$ (the only projection which is conserved in strongly deformed nuclei) and the parity $\pi$. 
In Ref.~\cite{Sal09} the structure of the intermediate 
$| 0^+\rangle$ and $| 1^+\rangle$ states was obtained within the QRPA to calculate $2\nu\beta\beta$-decay NME $M^{2\nu}$. Here, the approach of Ref.~\cite{Sal09} is straightforwardly extended to calculate all possible $| K^\pi\rangle$ states needed to construct the NME $M^{0\nu}$.

The intrinsic states $| K^\pi,m\rangle$ are generated within the
QRPA by a phonon creation operator acting on the ground-state wave function: 
\begin{equation}
| K^\pi,m\rangle=Q_{m,K}^\dagger |0^+_{g.s.}\rangle;~~~~Q_{m,K}^\dagger = \sum_{pn} X^{m}_{pn, K} A^\dagger _{pn, K} - Y^{m}_{pn, K} \bar{A}_{pn,K}.
\label{3}
\end{equation}
Here, $A^\dagger _{pn,K}=a^\dagger_{p}{a}^{\dagger}_{\bar{n}}$ and $\bar{A}_{pn,K}={a}_{\bar{p}}{a}_{n}$ 
are the two-quasiparticle creation and annihilation operators, 
respectively, with the bar denoting the time-reversal operation.
The quasiparticle pairs $p\bar n$ are defined by the selection rules $\Omega_p -\Omega_n = K$ 
and $\pi_p\pi_n=\pi$, where $\pi_\tau$ is the single-particle (s.p.) parity and  $\Omega_\tau$ is the projection 
of the total s.p. angular momentum on the nuclear symmetry axis ($\tau = p,n$).
The s.p. states $|p\rangle$ and $|n\rangle$ 
of protons and neutrons are calculated by solving the Schr\"odinger equation with the deformed axially symmetric Woods-Saxon potential~\cite{Sal09}. 
In the cylindrical coordinates the  
deformed Woods-Saxon s.p. wave functions $|\tau \Omega_\tau\rangle$ with $\Omega_\tau>0$ are decomposed over the deformed harmonic oscillator 
s.p. wave functions (with the principal quantum numbers $(N n_{z} \Lambda)$) and the spin wave functions 
$|\Sigma=\pm \frac12\rangle$:
 \begin{eqnarray}
|\tau \Omega_\tau\rangle &=& \sum_{N n_z \Sigma} b_{N n_z \Sigma} |N n_z \Lambda_\tau=\Omega_\tau-\Sigma\rangle |\Sigma\rangle,
\label{tau}
\end{eqnarray}
where $N=n_\perp+n_z$ ($n_\perp=2n_\rho+|\Lambda|$), $n_z$ and $n_\rho$ are the number of nodes
of the basis functions in the $z$- and $\rho$-directions,
respectively; $\Lambda =  \Omega - \Sigma$ and $\Sigma$ are the
projections of the orbital and spin angular momentum onto the symmetry axis $z$.
For the s.p. states with the negative projection $\Omega_\tau=-|\Omega_\tau|$, 
that are degenerate in energy with $\Omega_\tau=|\Omega_\tau|$,
the time-reversed version of Eq.~(\ref{tau}) is used as a definition (see also Ref.~\cite{Sal09}).
The states $(\tau,\bar \tau)$ comprise the whole single-particle model space.

The deformed harmonic oscillator wave functions $|N n_z \Lambda\rangle$  
can be further decomposed over the spherical harmonic oscillator ones   
$|n_rl\Lambda\rangle$ by calculating the corresponding spatial  
overlap integrals $A^{n_rl}_{N n_z \Lambda} =\langle n_r l \Lambda|N n_{z} \Lambda\rangle$  
($n_r$ is the radial quantum number,  $l$ and $\Lambda$ are the orbital angular momentum and its projection onto $z$-axes, respectively), see Appendix of Ref.~\cite{Sal09} for more details.   
Thereby, the wave function (\ref{tau}) can be reexpressed as   
\begin{eqnarray} 
|\tau \Omega_\tau\rangle&=& \sum_{\eta}B^{\tau}_{\eta}|\eta\Omega_\tau\rangle , 
\label{4} 
\end{eqnarray} 
where $|\eta\Omega_\tau\rangle=\sum\limits_{\Sigma} C^{j\Omega_\tau}_{l~ \Omega_\tau-\Sigma~\frac12~\Sigma} 
|n_r l \Lambda=\Omega_\tau-\Sigma\rangle |\Sigma\rangle$ is the spherical harmonic oscillator wave function in the $j$-coupled scheme ($\eta=(n_rlj)$), and  
$B^{\tau}_{\eta}= \sum\limits_{\Sigma}C^{j\Omega_{\tau}}_{l ~\Omega_{\tau}-\Sigma~\frac12~\Sigma}\, A^{n_r l}_{N n_z \Omega_{\tau}-\Sigma}\, b_{N n_z \Sigma}$, with $C^{j\Omega_{\tau}}_{l ~\Omega_{\tau}-\Sigma~\frac12~\Sigma}$ being the Clebsch-Gordan coefficient. 

The  QRPA equations:
\begin{eqnarray}
\left( \matrix{ {\cal A}(K) & {\cal B}(K) \cr
-{\cal B}(K) & -{\cal A}(K) }\right) 
~\left( \matrix{ X^m_K \cr Y^m_K} \right)~ = ~
\omega_{K,m}
~\left( \matrix{ X^m_K \cr Y^m_K} \right),
\label{5}
\end{eqnarray}
with realistic residual interaction are solved to get the forward $X^m_{i K}$,  
backward $Y^m_{i K}$ amplitudes and the excitation energies $\omega^{m_i}_K $ and $ \omega^{m_f}_K$  of  the $m$-th $K^\pi$ 
state in the intermediate nucleus. The matrix $\cal A$ and $\cal B$ are defined by
\begin{eqnarray}
{{\cal A}_{p n,{p'}{n'}}}(K)&=&{\delta}_{p n,{p'}{n'}}(E_p+E_n)+g_{pp}(u_p u_n u_{p'} u_{n'}+ v_p v_n v_{p'} v_{n'}) 
V_{p \bar{n}p'\bar{n'}}\nonumber\\
&&~~~~~~~~~~~~~~~~~~~~~~-g_{ph}(u_p v_n u_{p'} v_{n'}+ v_p u_n v_{p'} u_{n'})
V_{p n'p'n}\nonumber\\
{{\cal B}_{p n,{p'}{n'}}}(K)&=&-g_{pp}(u_p u_n v_{p'} v_{n'}+ v_p v_n u_{p'} u_{n'}) 
V_{p \bar{n}p'\bar{n'}}\nonumber\\
&&-g_{ph}(u_p v_n v_{p'} v_{n'}+ v_p u_n u_{p'} v_{n'})
V_{p n'p'n}
\label{6}
\end{eqnarray}  
where $E_p+E_n$ are the two-quasiparticle excitation energies, 
$V_{p n,{p'}{n'}}$ and $V_{p \bar {n},{p'}\bar {n'}}$ are the $p-h$ and $p-p$
matrix elements of the residual nucleon-nucleon interaction $V$, respectively,
$u_\tau$ and $v_\tau$ are the coefficients of the Bogoliubov transformation.

As a residual two-body interaction we use the nuclear Brueckner $G$~matrix, that is a solution of the Bethe-Goldstone equation, derived from the Bonn-CD one boson exchange potential, as used also in the spherical calculations of Ref.~\cite{Rod05}. 
The $G$~matrix elements are originally calculated with respect to a spherical harmonic oscillator s.p. basis.
By using the decomposition of the deformed s.p. wave function in Eq.~(\ref{4}), 
the two-body deformed wave function 
can be represented as:
\begin{eqnarray}
|p \bar{n}\rangle&=& \sum_{\eta_p\eta_n J}
F^{JK}_{p\eta_p n\eta_n}|\eta_p \eta_n,J K\rangle,
\label{decomp}
\end{eqnarray}
where 
$|\eta_p \eta_n, J K \rangle=\sum_{m_{p} m_{n}}
C^{JK}_{j_p m_{p} j_n m_{n} }|\eta_p m_{p}\rangle|\eta_n m_{n}\rangle$,
and  
$F^{JK}_{p\eta_p n\eta_n}= B^p_{\eta_p}B^n_{\eta_n}(-1)^{j_n-\Omega_{n}}C^{JK}_{j_p\Omega_{p} j_n-\Omega_{n}}$ 
is defined for the sake of simplicity
($(-1)^{j_n-\Omega_{n}}$ is the phase arising from the time-reversed states $|\bar{n}\rangle$).
The particle-particle $V_{p \bar {n},~{p'}\bar {n'}}$ and particle-hole $V_{p n',~p'n}$ 
interaction matrix elements in the representation (\ref{6}) 
for the QRPA matrices ${\cal A,\ B}$ (\ref{5}) in the deformed Woods-Saxon 
single-particle basis can then be given in terms of the spherical $G$~matrix elements as follows:
\begin{eqnarray}
V_{p \bar {n},~{p'}\bar {n'}}&= -&
2\sum_{J}\sum_{{\eta}_{p}{\eta}_{n}} \sum_{{\eta}_{p'}{\eta}_{n'}}
F^{JK}_{p\eta_p n\eta_n}F^{JK}_{p'\eta_{p'} {n'}\eta_{n'}}
G(\eta_p\eta_n\eta_{p'}\eta_{n'},J),\\
V_{p n',~p'n}&=&
2\sum_{J}\sum_{{\eta}_{p}{\eta}_{n}} \sum_{{\eta}_{p'}{\eta}_{n'}}
F^{JK'_{pn'}}_{p\eta_p {\bar n}'\eta_{n'}}
F^{JK'_{pn'}}_{p'\eta_{p'} \bar{n}\eta_{n}} G(\eta_p\eta_{n'}\eta_{p'}\eta_{n},J),
\label{10}
\end{eqnarray}
where $K'_{pn'}=\Omega_p+\Omega_{n'}=\Omega_{p'}+\Omega_n$.

The matrix element $M^{0\nu}$ is given within the QRPA in the intrinsic system by a sum of the partial amplitudes of transitions via all the intermediate states $K^\pi$:
\begin{equation}
M^{0\nu}=\sum_{K^\pi} M^{0\nu}(K^\pi)\ ; \ M^{0\nu}(K^\pi) = 
\sum_{\alpha} s^{(def)}_\alpha O_\alpha(K^\pi). \label{M0n}
\end{equation}
Here, we use the notation of Appendix B in Ref.~\cite{anatomy}, $\alpha$ stands for the set of four single-particle indices $\{p,p',n,n'\}$, 
and $O_\alpha(K^\pi)$ is a two-nucleon transition amplitude via the $K^\pi$ states
in the intrinsic frame:
\begin{equation}
O_\alpha(K^\pi)=\sum_{m_i,m_f}
\langle 0_f^+|c_{p}^\dagger c_{n}|K^\pi m_f\rangle 
\langle K^\pi m_f|K^\pi m_i\rangle
\langle K^\pi m_i|c^\dagger_{p'} c_{n'}|0_i^+\rangle .
\label{O}
\end{equation}
The two sets of intermediate nuclear states generated from the
initial and final g.s. (labeled by $m_i$ and $m_f$, respectively) 
do not come out identical within the
QRPA. A standard way to tackle this problem is to introduce in Eq.~(\ref{O}) the overlap factor of these states $\langle K^\pi m_f|K^\pi m_i\rangle$, whose representation is given below, Eq.~(\ref{overlap}).
Two-body matrix elements $s^{(def)}_\alpha$ of the neutrino potential in 
Eq.~(\ref{M0n}) in a deformed Woods-Saxon single-particle basis are decomposed over the the spherical harmonic oscillator ones according to 
Eqs.~(\ref{decomp},\ref{10}):
\begin{equation}
s^{(def)}_{pp'nn'}=
\sum_{J}\sum_{\footnotesize\begin{array}{c}
\eta_p \eta_{p'}\\[-1pt]  \eta_n \eta_{n'}\end{array}}
F^{JK}_{p\eta_p n\eta_n}F^{JK}_{p'\eta_{p'}n'\eta_{n'}}s^{(sph)}_{\eta_p\eta_{p'} \eta_n\eta_{n'}}(J),
\end{equation}
\begin{eqnarray}
s^{(sph)}_{pp'nn'}(J)&=&\displaystyle \sum_{\mathcal J}
(-1)^{j_n + j_{p'} + J + {\mathcal J}} \hat{\mathcal J}
\left\{
\begin{array}{c c c}
j_p & j_n & J \\ j_{n'} & j_{p'} & {\mathcal J}
\end{array}
\right\} 
\langle p(1), p'(2); {\mathcal J} \| {\mathcal O_\ell}(1,2) \| n(1), n'(2); {\mathcal J} \rangle\,,
\end{eqnarray}
where $\hat{\mathcal{J}} \equiv \sqrt{2\mathcal{J}+1}$, and ${\mathcal O_\ell}(1,2)$ is the neutrino potential as a function of coordinates of two particles, with ${\ell}$ labeling its Fermi (F), Gamow-Teller (GT) and Tensor (T) parts. 

The particle-hole transition amplitudes in Eq.~(\ref{O}) can be represented in terms 
of the QRPA forward $X^m_{i K}$ and backward $Y^m_{i K}$  amplitudes along with 
the coefficients of the Bogoliubov transformation $u_\tau$ and $v_\tau$~\cite{Sal09}:
\begin{eqnarray}
\langle 0_f^+|c_{p}^\dagger c_{n}|K^\pi m_f\rangle&=&v_{p}u_{n}X^{m_f}_{pn,K^\pi}+u_{p}v_{n}Y^{m_f}_{pn,K^\pi},\nonumber\\
\langle K^\pi m_i|c^\dagger_p c_{n}|0_i^+\rangle&=&u_{p}v_{n}X^{m_i}_{pn,K^\pi}+v_{p}u_{n}Y^{m_i}_{pn,K^\pi}.\nonumber
\end{eqnarray}
The overlap factor in Eq.~(\ref{O}) can be written as:
\begin{eqnarray}
\langle K^\pi m_f|K^\pi m_i\rangle&=&\sum_{l_il_f}[X^{m_f}_{l_fK^\pi}X^{m_i}_{l_iK^\pi}-Y^{m_f}_{l_fK^\pi}Y^{m_i}_{l_iK^\pi}]
\mathcal{R}_{l_fl_i}\langle BCS_f|BCS_i\rangle
\label{overlap}
\end{eqnarray}
Representations for ${\cal R}_{l_fl_i}$  and the overlap factor $\langle BCS_f|BCS_i\rangle$ between the initial and final BCS vacua  are given in Ref.~\cite{Sim03}.

\section{Results and Analysis}

We have computed the NME $M^{0\nu}$ for the $0\nu\beta\beta$ decays $^{76}$Ge$\rightarrow ^{76}$Se, $^{150}$Nd$\rightarrow ^{150}$Sm, $^{160}$Gd$\rightarrow ^{160}$Dy.

The single-particle Schr\"odinger equation with the Hamiltonian of 
a deformed Woods-Saxon mean field is solved on the basis of a axially-deformed 
harmonic oscillator. The parametrization of the mean field is adopted from the spherical
calculations of Refs.~\cite{Rod05,anatomy,Sim09}.  
We use here the single-particle deformed basis corresponding in the spherical limit to full (4--6)$\hbar\omega$ shells.
Decomposition of the deformed single-particle wave functions is performed over the spherical harmonic oscillator states within the seven major shells.
Only quadrupole deformation is taken into account in the calculation. The geometrical quadrupole deformation parameter $\beta_2$ of the deformed Woods-Saxon mean 
field is obtained 
by fitting the experimental deformation parameter $\beta= \sqrt{\frac{\pi}{5}}\frac{Q_p}{Z r^{2}_{c}}$, where $r_c $ is the charge rms radius and  $Q_p$ is the empirical intrinsic quadrupole moment. The latter 
can be derived from the laboratory quadrupole moments measured by the Coulomb excitation reorientation technique, or from the corresponding 
B(E2) values~\cite{ragha}. 
We take in this work experimental values 
extracted from the B(E2) values as being more accurate.
The fitted values of the parameter $\beta_2$ of the deformed Woods-Saxon mean 
field, which allow us to reproduce the experimental $\beta$, are listed in Table~\ref{table.1}. We label these sets of parameters as ``1". The spherical limit, i.e. $\beta_2=0$, is considered as well (labeled as ``0"), to compare with the earlier results of Ref.~\cite{Rod05}. The adopted here procedure of fitting $\beta_2$ is more consistent than the approximate ansatz $\beta_2=\beta$ used in Ref.~\cite{Sal09}.

As in Refs.~\cite{Rod05,Sal09,anatomy,Sim09,Fang10}, the nuclear Brueckner $G$~matrix, obtained by a solution of the Bethe-Goldstone equation with the Bonn-CD one boson exchange nucleon-nucleon potential, is used as a residual two-body interaction. 
First, the BCS equations are solved to obtain the Bogoliubov coefficients, gap parameter and the chemical potentials. To solve the QRPA equations,
one has to fix the particle-hole $g_{ph}$ and particle-particle
$g_{pp}$ renormalization factors of the residual interaction, Eqs.~\ref{6}. As in  Ref.~\cite{Sal09,Fang10}, we determine a value of $g_{ph}$ 
by fitting the experimental  position of the Gamow-Teller giant resonance (GTR) in the intermediate nucleus. Since there is no experimental information on the GTR energy for $^{150}$Nd, we use for this nucleus the same $g_{ph}=0.90$ as fitted for $^{76}$Ge (this value is slightly different from the fitted $g_{ph}=1.15$ of Ref.~\cite{Sal09} 
because of a different parametrization of 
the mean field used here). 
The parameter $g_{pp}$ can be determined by fitting the experimental value of the $2\nu\beta\beta$-decay NME $M^{2\nu}_{GT}=0.07$ MeV$^{-1}$~\cite{Bar10}.
To account for the quenching of the axial-vector coupling constant $g_A$, we choose in the calculation, along with the bare value $g_A=1.25$, also the quenched value $g_A^{qch}=0.75\cdot g_A=0.94$, where the quenching factor of 0.75 comes from 
a recent experimental measurement of GT strength distribution in $^{150}$Nd~\cite{Zeg09}.
The two sets of the fitted values of $g_{pp}$ corresponding to the cases without or with quenching of $g_A$ are listed Table~\ref{table.1} as cases (I) and (II), respectively. 
Note, that the more realistic procedure of fitting $\beta_2$ adopted here also give us more realistic $g_{pp}\simeq 1$ values as compared with those of Ref.~\cite{Sal09}.

\begin{table}[h]
\centering
\caption{The values of the deformation parameter of Woods-Saxon mean field 
$\beta_2$ for initial
 (final) nuclei fitted in the calculation to reproduce the experimental quadrupole moment (labeled as ``1"). 
The spherical limit is labeled as ``0".
Also the fitted values of the particle-particle strength parameter $g_{pp}$ are listed 
(for both cases without~(I) and with quenching~(II) of $g_A$). 
The particle-hole strength parameter is $g_{ph}=0.90$.
The BCS overlap factor $\langle BCS_f|BCS_i\rangle$ (\ref{overlap}) between the initial and final BCS vacua is given in the last column.}
\label{table.1}
\begin{tabular}{|l|c c|c|c|c|}
	\hline
 & & & & & \\
initial (final) &  $\beta_{2}$ & & $g_{pp}$ (I) &$g_{pp}$ (II) & $\langle BCS_i|BSC_f\rangle$\\	
nucleus & & & & & \\[4pt]	

\hline
$^{76}$Ge ($^{76}$Se)& 0.10 (0.16) & ``1" & 0.71 & 0.66 & 0.74 \\
                     & 0.0\ \ (0.0) & ``0" & 0.68 & 0.63 & 0.81 \\ 
\hline
$^{150}$Nd ($^{150}$Sm)& 0.240 (0.153) & ``1" & 1.05 & 1.00 & 0.52 \\
                       &  0.0\ \ (0.0) & ``0" & 1.01 & 0.99 & 0.85 \\
\hline
$^{160}$Gd ($^{160}$Dy)& 0.303 (0.292) & ``1" & 1.00\footnote{As there is no experimental value of $M^{2\nu}$ for $^{160}$Gd, we do not renormalize the p-p interaction and use $g_{pp}=1$.} 
& 1.00 & 0.74 \\ 
\hline 
\end{tabular}
\end{table}

Having solved the QRPA equations, the two-nucleon transition amplitudes (\ref{O}) are calculated and, by combining them with the two-body matrix elements of the neutrino potential, the total $0\nu\beta\beta$ NME $M^{0\nu}$ (\ref{M0n}) is formed. The present computation is rather time consuming since numerous programming loops are needed to calculate the decompositions of the two-body matrix elements in the deformed basis over the spherical ones. Therefore, to speed up the calculations the mean energy of 7 MeV of the intermediate states is used in the neutrino propagator.
Following Refs.~\cite{Rod05,anatomy,Sim09}, we have taken into account
the effects of the finite nucleon size and higher-order weak currents are included.
Recently, it was shown~\cite{Sim09} that a modern self-consistent treatment of the two-nucleon short-range correlations (s.r.c.) leads to a change in the NME $M^{0\nu}$  only by a few percents, much less than the traditional Jastrow-type representation of the s.r.c. does. A very similar effect is found in the present calculation (see below).

We start our discussion of the calculated $0\nu\beta\beta$-decay NME 
by a comparison of the matrix elements of this work obtained in the spherical limit 
with the previous ones of Refs.~\cite{Rod05,anatomy,Sim09}, that provides an important 
cross-check of the present calculation. Though formally the adiabatic Bohr-Mottelson approximation is not applicable in the limit of vanishing deformation,
it is easy to see that the basic equations~(\ref{M0n})--(\ref{overlap}) do have the correct spherical limit. 

According to Eq.~(\ref{M0n}), 
the total calculated $0\nu\beta\beta$-decay NME is formed by the sum of the all 
partial contributions $M^{0\nu}(K^\pi)$ of different intrinsic intermediate states $K^\pi$, with $M^{0\nu}(|K|^\pi)=M^{0\nu}(-|K|^\pi)$.
In the spherical limit the total angular momentum $J$ becomes a good quantum number, and the intermediate $K^\pi$ states corresponding to a given $J^\pi$ state become degenerate. In addition, in this limit each projection $K$ of an intermediate $J^\pi$ state contributes equally to the calculated $0\nu\beta\beta$-decay NME, as a consequence of rotational symmetry.
To represent standard spherical results in the terms of the present paper,
one has to use the following expression for the spherical partial contribution of a projection $K$: $M^{0\nu}(K^\pi)=\sum\limits_{J\ge |K|}M^{0\nu}(J^\pi)/(2J+1)$ (it is easy to see that having summed over all $K^\pi$ one obtains the total NME 
$M^{0\nu}=\sum\limits_{J}M^{0\nu}(J^\pi)$). 
From this representation it can generally be expected that the smaller is $|K|$, the larger the corresponding partial contribution $M^{0\nu}(K^\pi)$ should be (since simply more $J$'s contribute, and their contributions are of the same sign in most cases, see~\cite{Rod05}). 
This behavior is in fact revealed by most of the calculation results (see below).

To test our new numerical code calculating $0\nu\beta\beta$-decay NME
for deformed nuclei, we have taken the spherical limit and used in it the spherical
harmonic oscillator wave functions as usually done in the QRPA calculations~\cite{Rod05,anatomy,Sim09}. 
The NME calculated by different codes are found to be in an excellent agreement.

\begin{figure}
\includegraphics[scale=0.35]{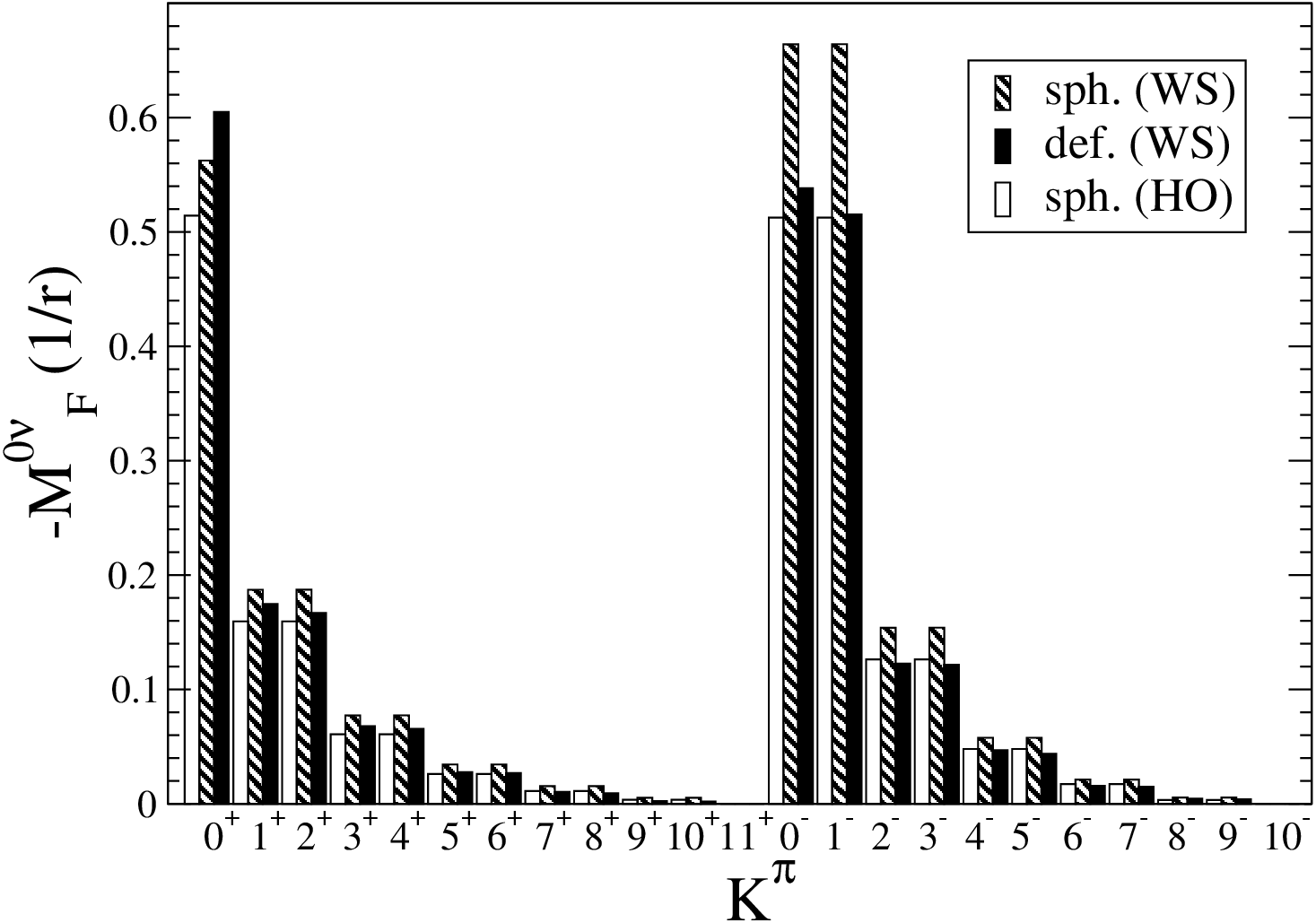}\includegraphics[scale=0.35]{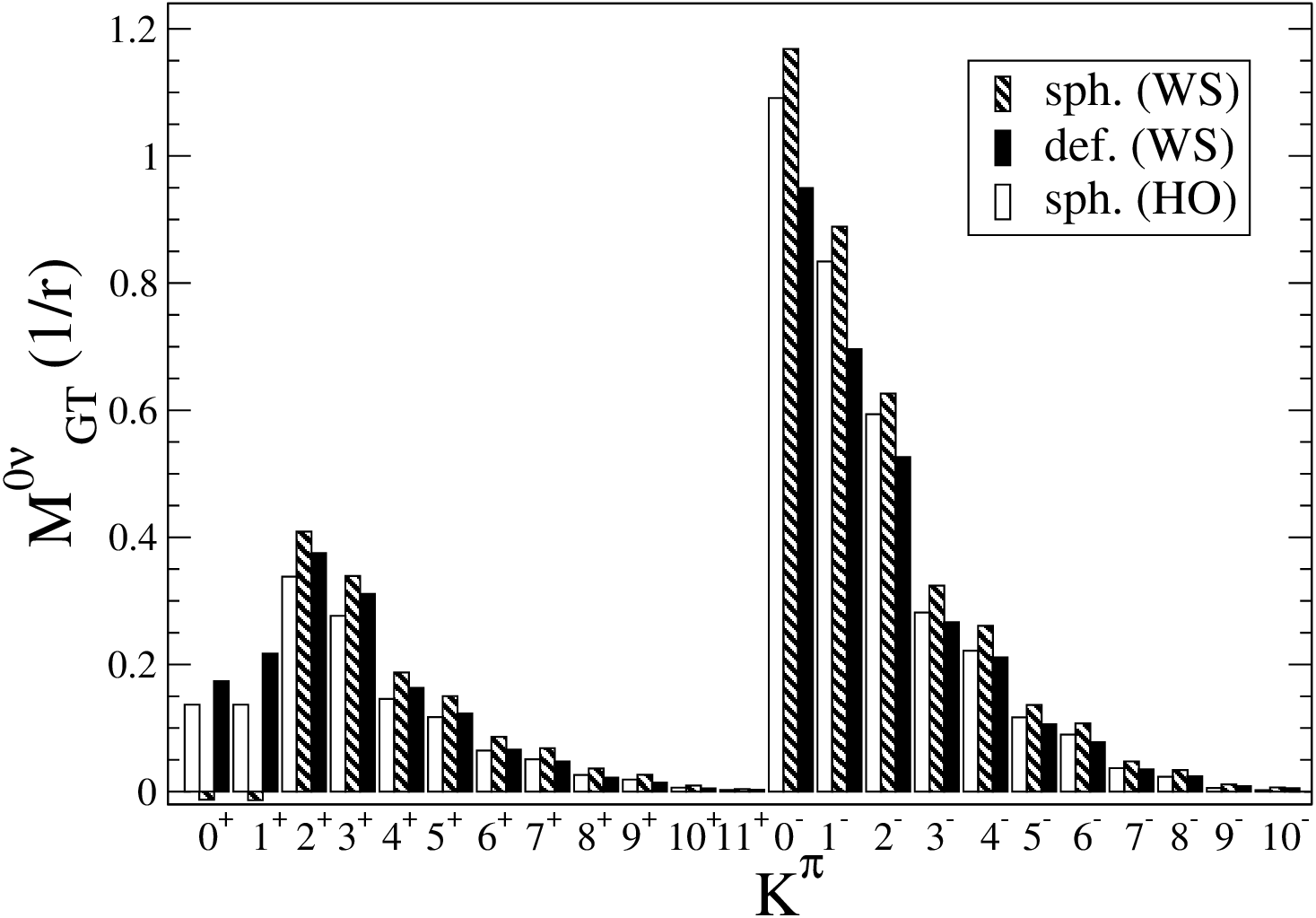}
\caption{The partial contributions $M^{0\nu}(K^\pi)$ of different intermediate $K^\pi$ states to $M^{0\nu}(1/r)$ for $^{150}$Nd$\rightarrow ^{150}$Sm in the cases of vanishing and realistic deformations. 
For simplicity, the BCS overlap factor is omitted in these results.
The Fermi $M^{0\nu}_F(1/r)$ and the GT $M^{0\nu}_{GT}(1/r)$ contributions are shown in the left and right panels, respectively.
The three bars represent (from left to right) the results obtained with the spherical
harmonic oscillator wave functions, with the Woods-Saxon wave functions in the spherical limit, and with the deformed Woods-Saxon wave functions for realistic deformations from Table~\ref{table.1}.}
\label{fig.1}
\end{figure}

For the further discussion we define the following contributions to the total $0\nu\beta\beta$-decay NME: 
$M^{0\nu}_F(1/r)$ and $M^{0\nu}_{GT}(1/r)$ are calculated by taking into account only the Coulomb-like radial dependence of the neutrino potential. The total corrections $\Delta M^{0\nu}_F$ and $\Delta M^{0\nu}_{GT}$ to $M^{0\nu}_F(1/r)$ and $M^{0\nu}_{GT}(1/r)$, respectively, come from the effects of the finite nucleon size (FNS), the closure energy, and higher order weak currents. 
Finally, the corrections $\delta_{i} M^{0\nu}_F$ and $\delta_{i} M^{0\nu}_{GT}$, respectively, come from the effects of the s.r.c. (Jastow-like s.r.c. is denoted by the subscript $i=1$, and the self-consistent Bonn-CD s.r.c. --- by the subscript $i=2$). Thus, the final total $0\nu\beta\beta$-decay NME is given by 
\be
M^{0\nu}=M^{0\nu}(1/r)+\Delta M^{0\nu}+\delta_i M^{0\nu}.\nonumber
\label{M00}
\ee
The effect of deformation and different choices of the single-particle wave functions on the partial contributions $M^{0\nu}(K^\pi)$ of different $K^\pi$ intermediate states to $M^{0\nu}(1/r)$ for $^{150}$Nd is illustrated in Fig.~\ref{fig.1}. 
The BCS overlap factor is neglected here for simplicity.
The Fermi and the GT contributions are shown in the left and right panels of the figure, respectively.
The left and middle bars for each $K^\pi$ represent the results obtained 
for zero deformation with the spherical harmonic oscillator wave functions and with the Woods-Saxon wave functions, respectively, and the right bar represents the result calculated with the deformed Woods-Saxon wave functions for the finite deformations from Table~\ref{table.1} (case ``1"). Each bar for $K>0$ represents a sum of the equal  contributions of the positive and negative projections, $M^{0\nu}(K^\pi)+M^{0\nu}(-K^\pi)=2M^{0\nu}(K^\pi)$.
Because the partial contributions $M^{0\nu}_F(K^\pi)$ to the Fermi NME in the spherical limit are non-zero only for the intermediate states of  natural parity $\pi=(-1)^J$, the equality   $M^{0\nu}_F(K^\pi,\pi=(-1)^K)=M^{0\nu}_F((K-1)^\pi,\pi=(-1)^K)$ ($K>0$) should hold in this limit (this simply reflects equality of the contributions of different projections $K$ for a given $J$). This equality is nicely fulfilled in our calculations, as illustrated by Fig.~\ref{fig.1}.
Also, one can see in Fig.~\ref{fig.1} that even for rather large deformations the  partial contributions $M^{0\nu}(K^\pi)$ (calculated with neglect of the BCS overlap) do not differ much from the corresponding ones of the spherical calculations.
The corresponding results for the other nuclei show very similar pattern.

\begin{figure}
\includegraphics[scale=0.33]{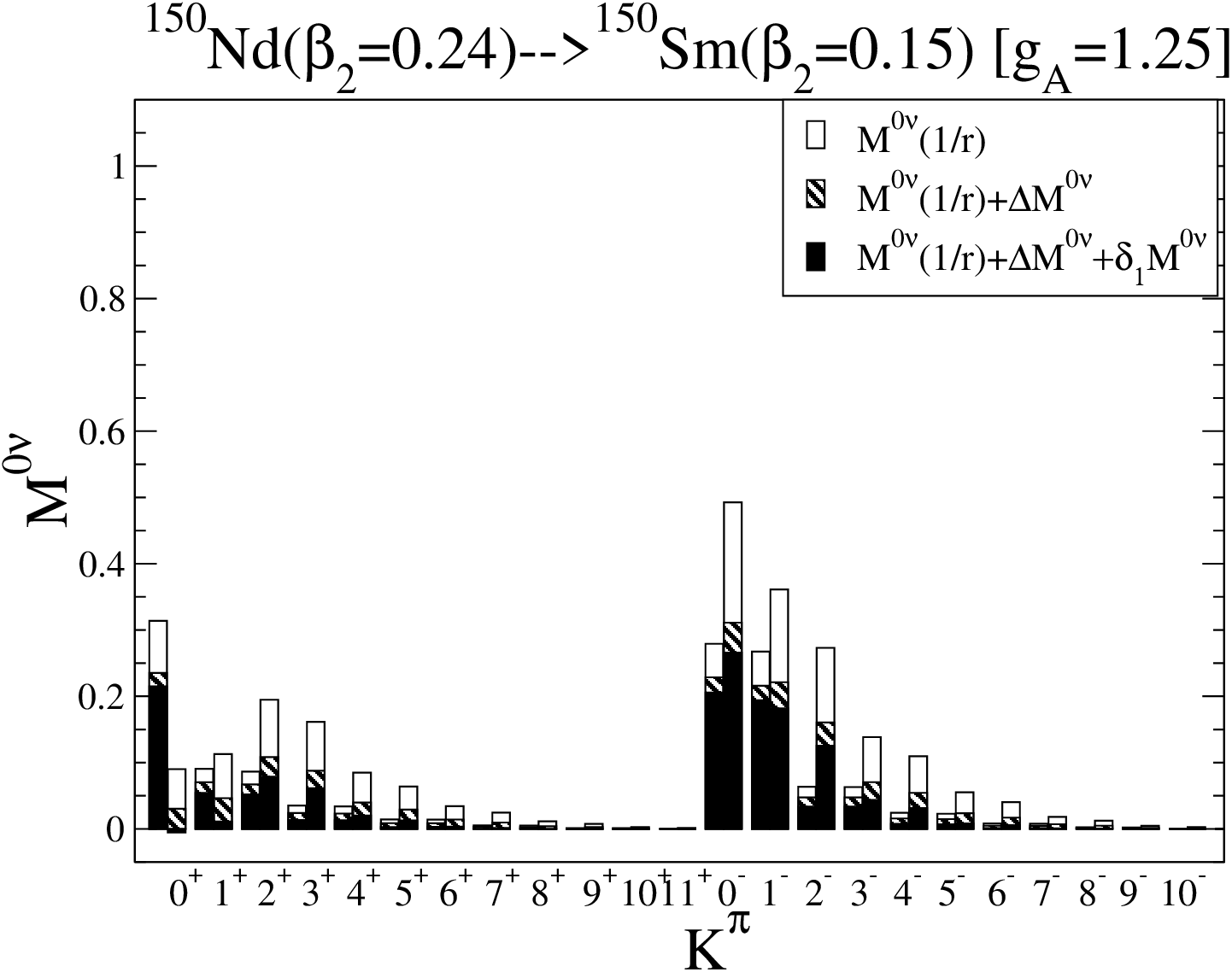}~~\includegraphics[scale=0.33]{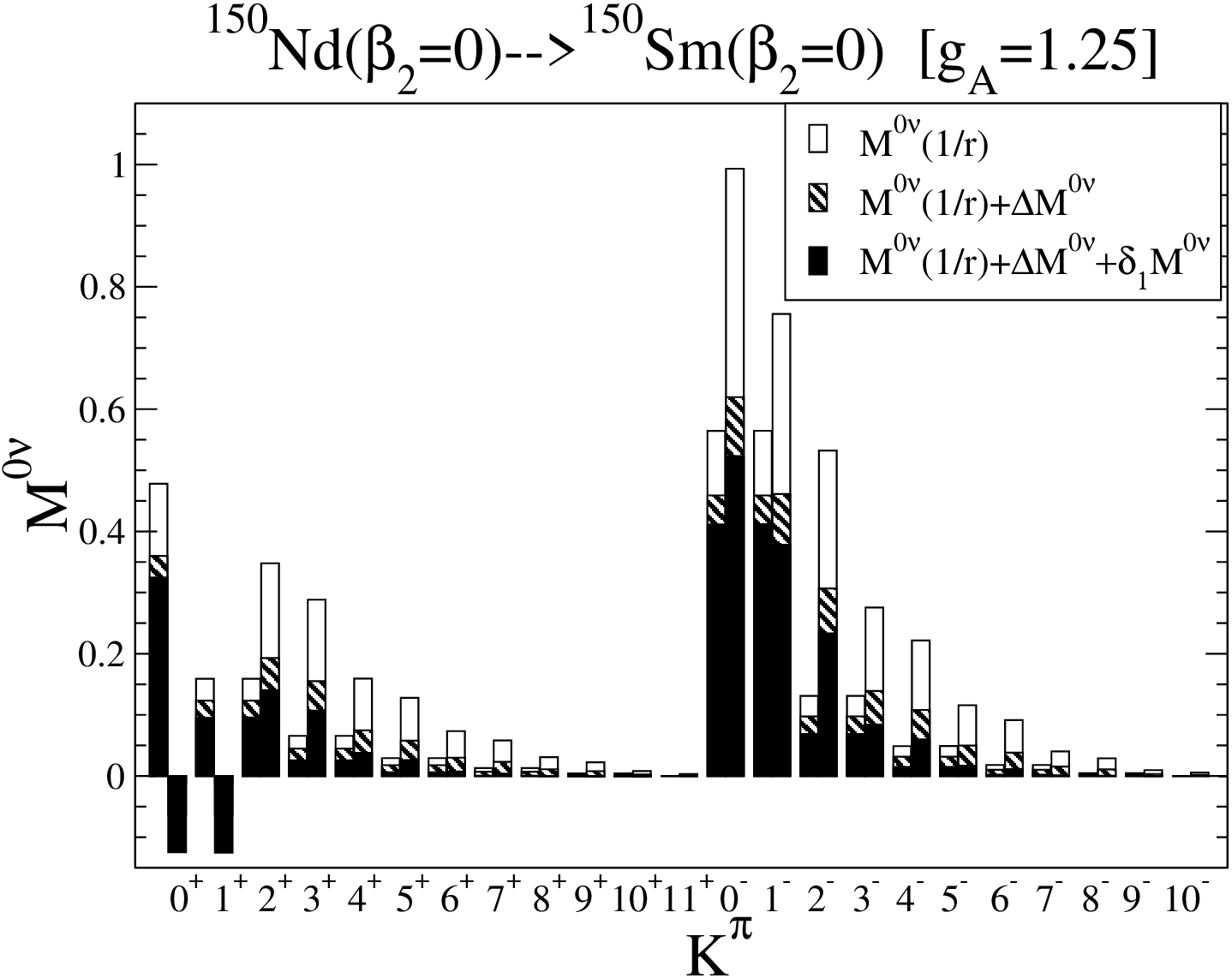}

\

\includegraphics[scale=0.33]{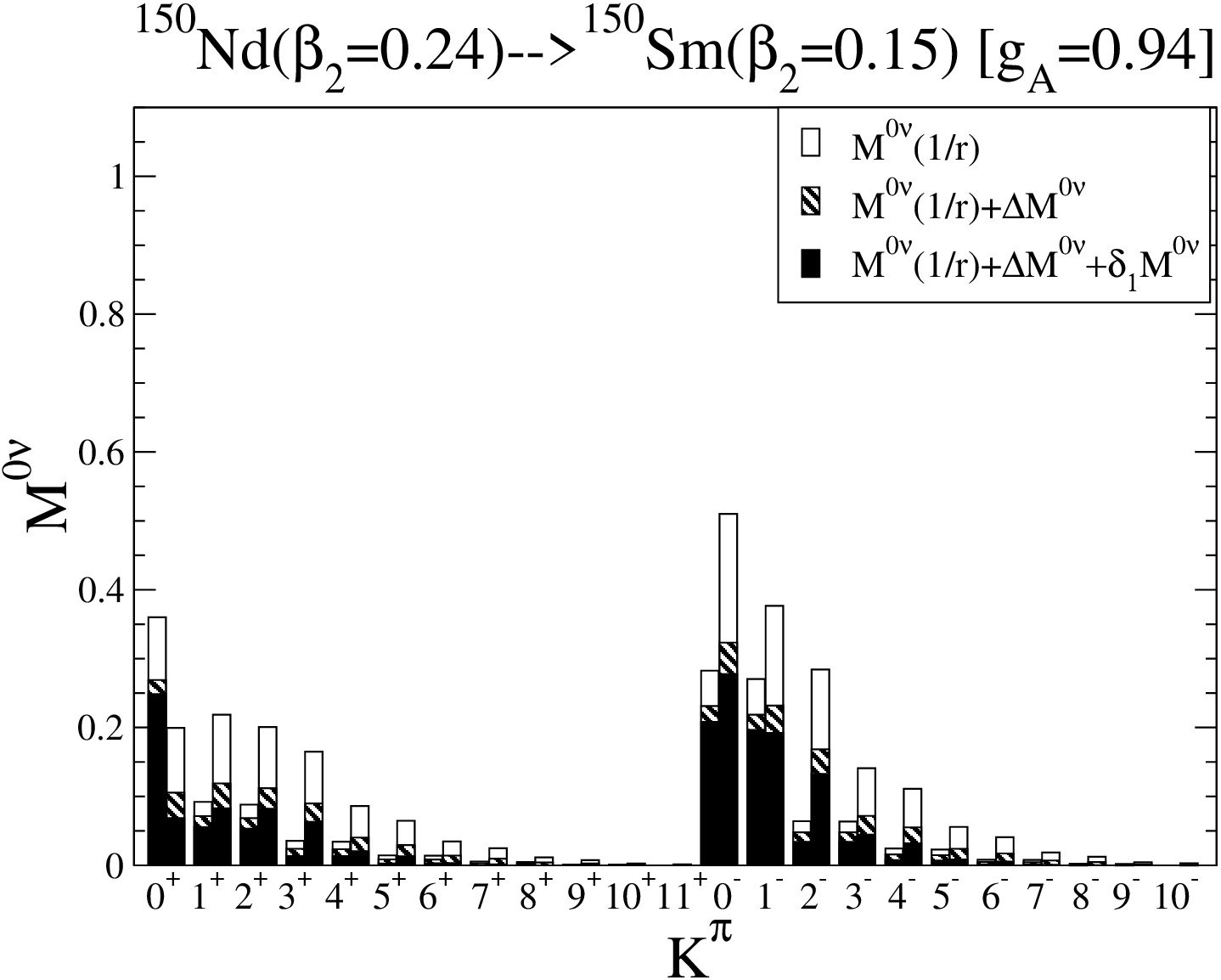}~~\includegraphics[scale=0.33]{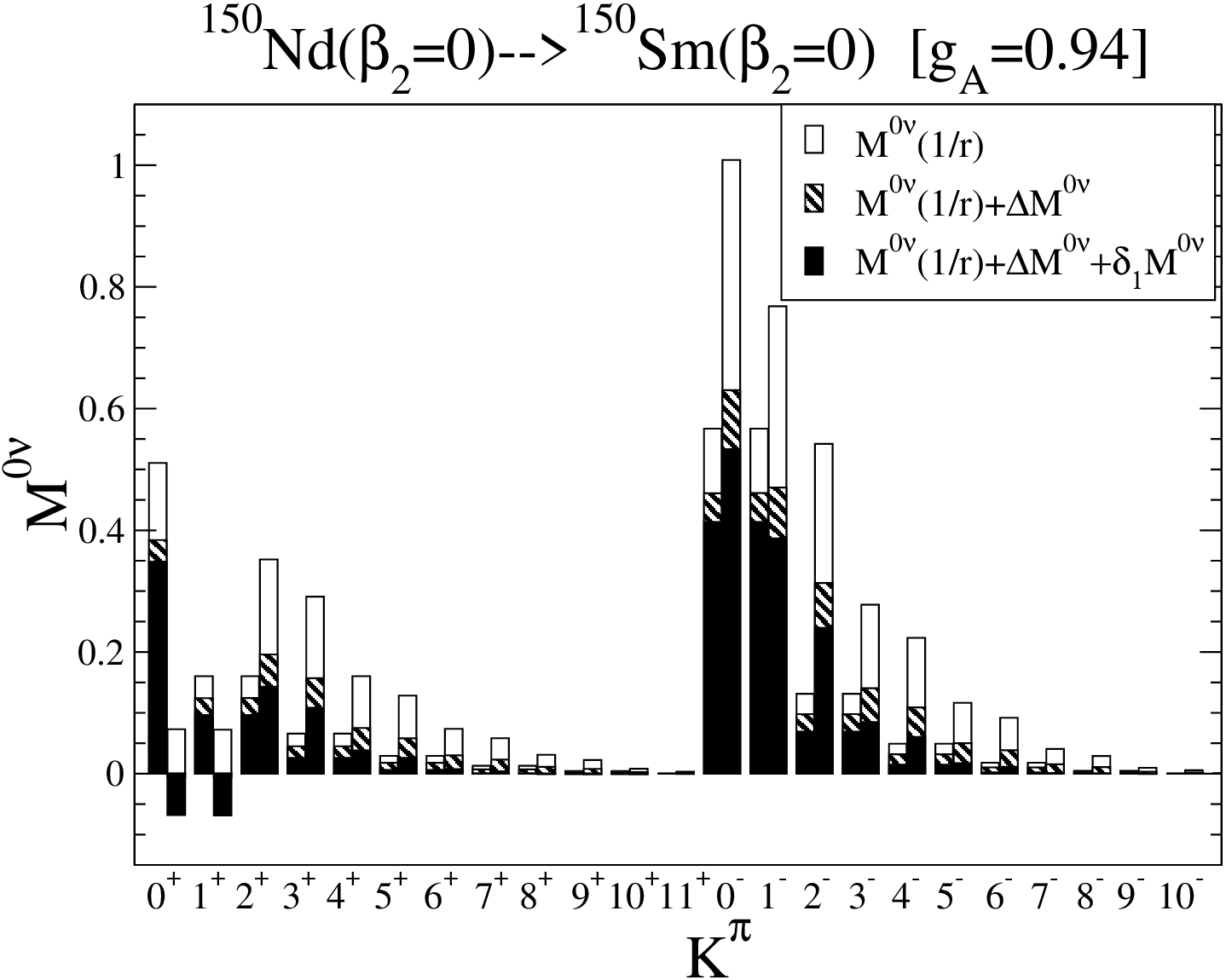}
\caption{ The partial contributions $M^{0\nu}(K^\pi)$ of different intrinsic intermediate states to the total $M^{0\nu}$ for $^{150}$Nd$\rightarrow ^{150}$Sm in the case of realistic deformation (left panels) and in the spherical limit (right panels). The Fermi $-M^{0\nu}_F(K^\pi)$ and GT $M^{0\nu}_{GT}(K^\pi)$ contributions are shown for each $K^\pi$ by the left and right bars, respectively.
The contributions of the leading, Coulomb-like, radial dependence of the neutrino potential is represented by the open bar (``$M(1/r)$"), with the effects of the FNS, the closure energy, and higher order weak currents included - by the hatched bar (``$M(1/r)+\Delta M$"), and the final contributions, including in addition the effect of the Jastow-like s.r.c., is represented by the filled bar (``$M(1/r)+\Delta M + \delta_1 M$"). 
The upper and lower panels show the results corresponding to the unquenched $g_A=1.25$ and quenched $g_A=0.94$ (fitted values (I) and (II) of $g_{pp}$, respectively, see Table~\ref{table.1}). 
}
\label{fig.2}
\end{figure}

\begin{figure}
\includegraphics[scale=0.33]{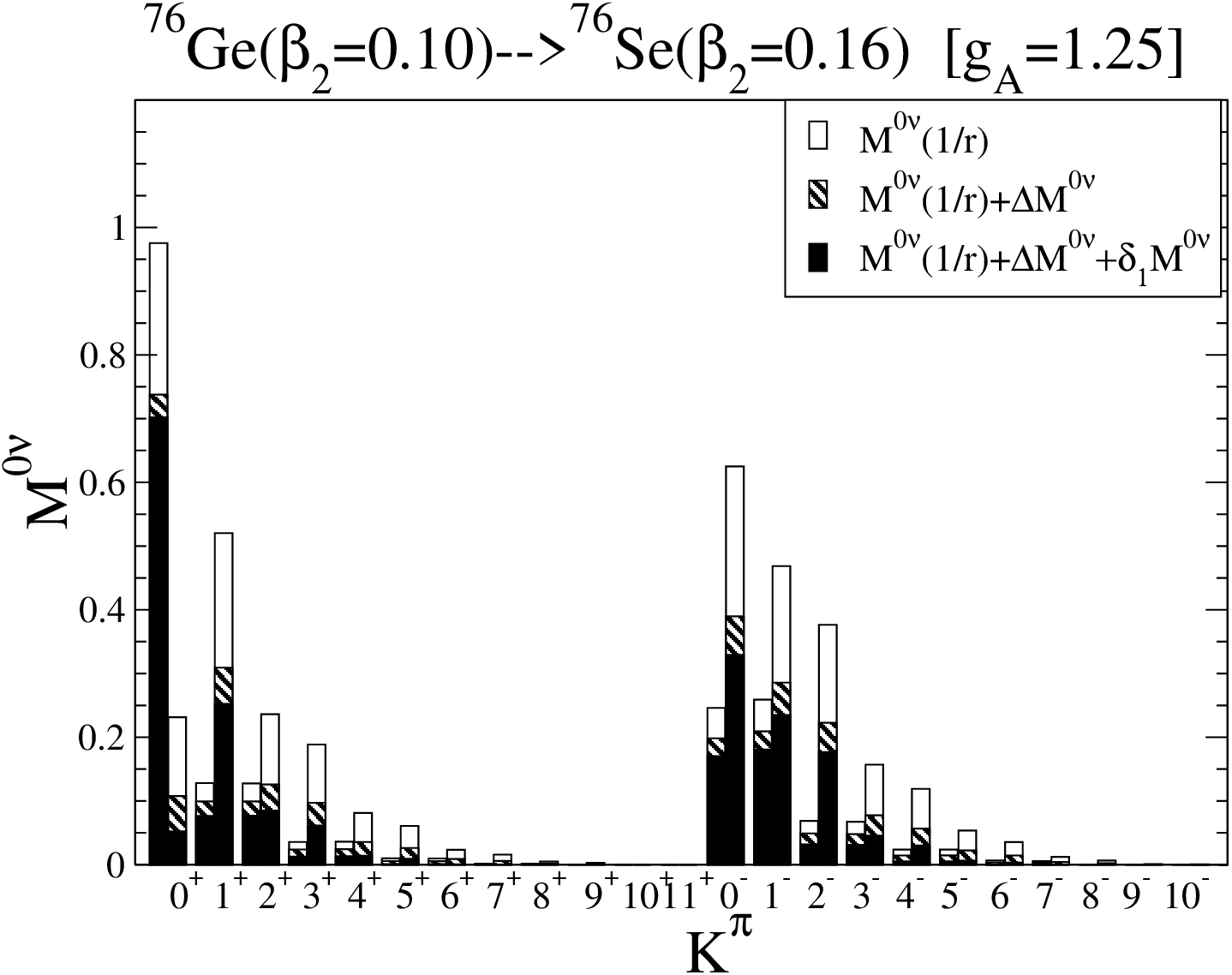}~~\includegraphics[scale=0.33]{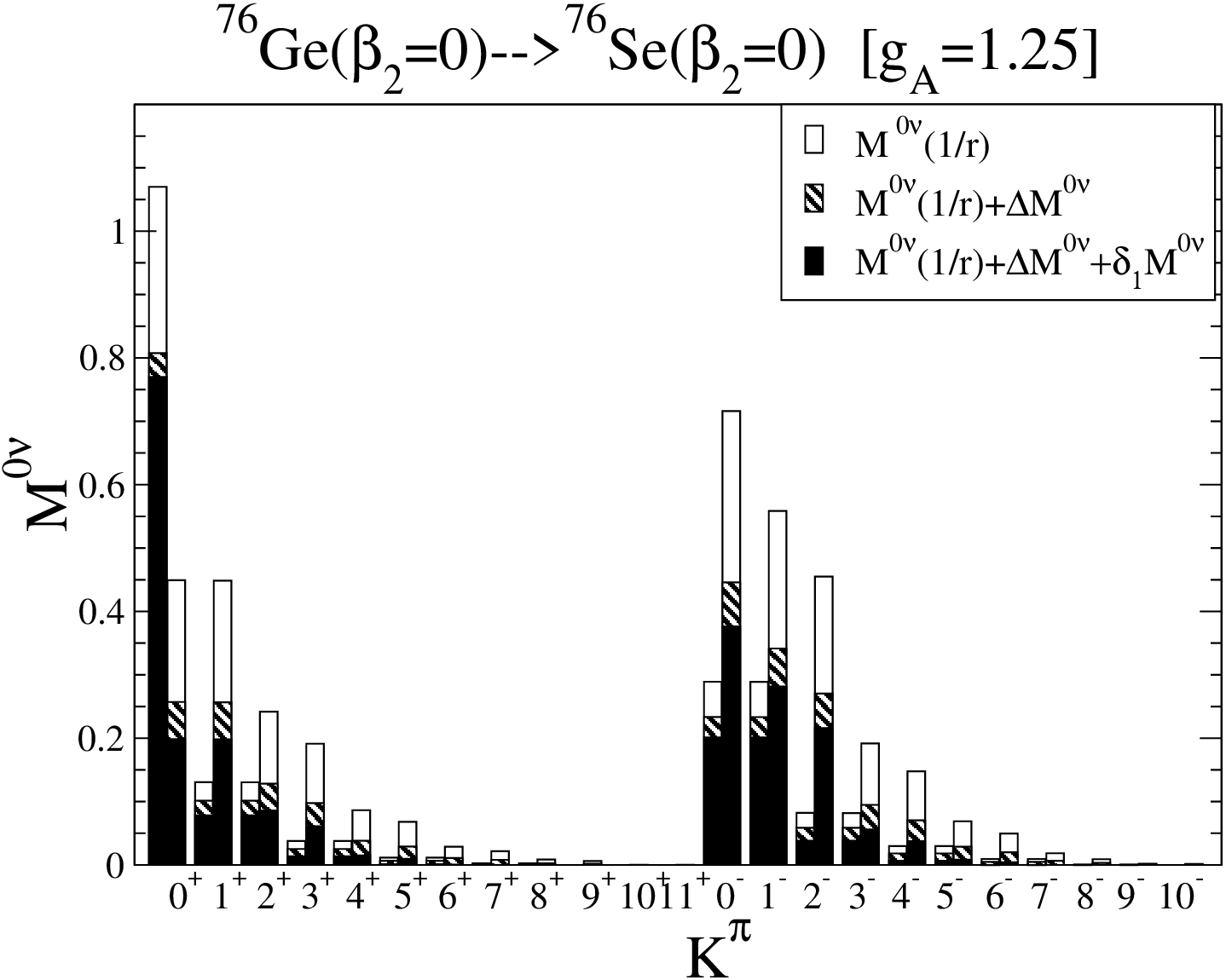}

\

\includegraphics[scale=0.33]{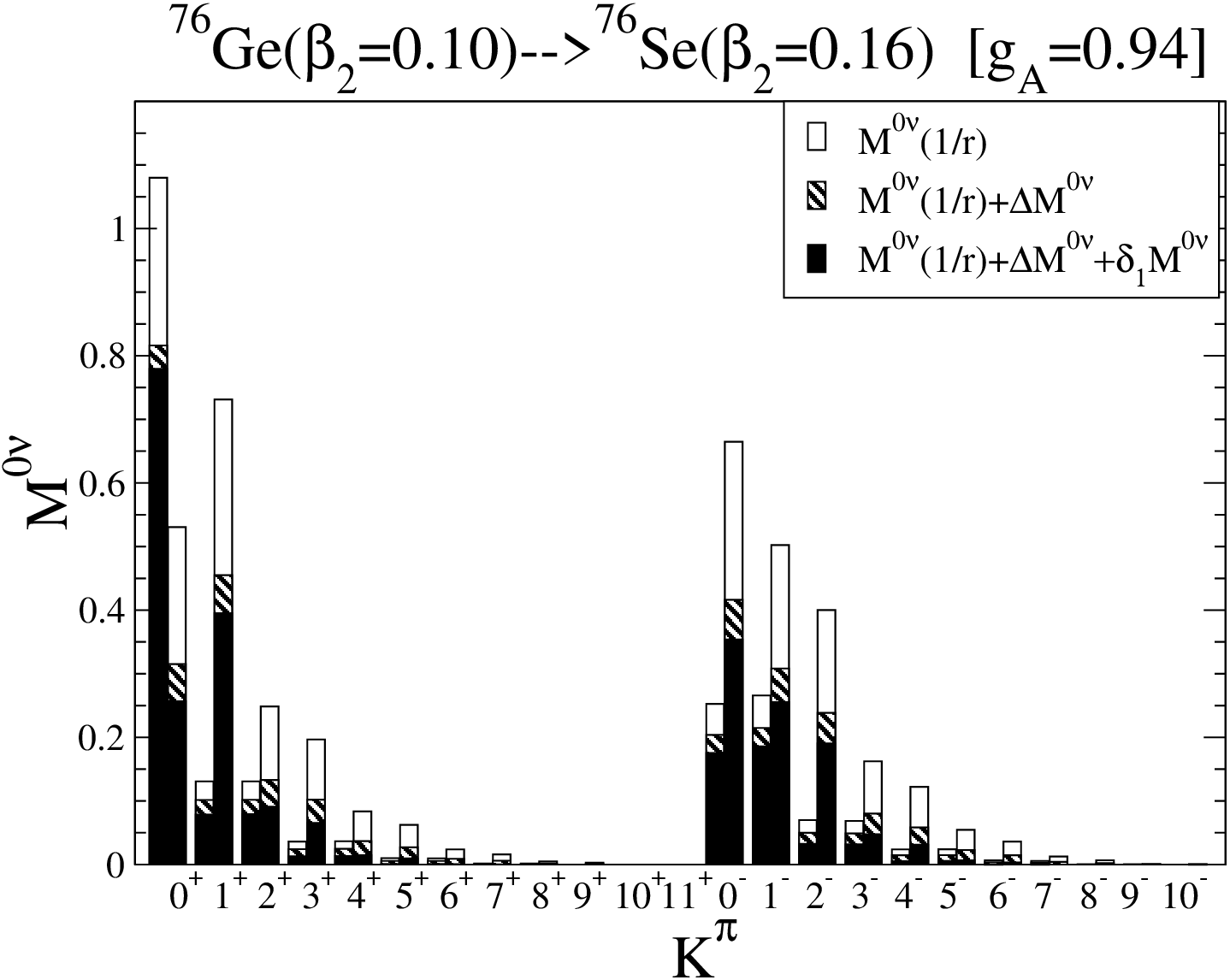}~~\includegraphics[scale=0.33]{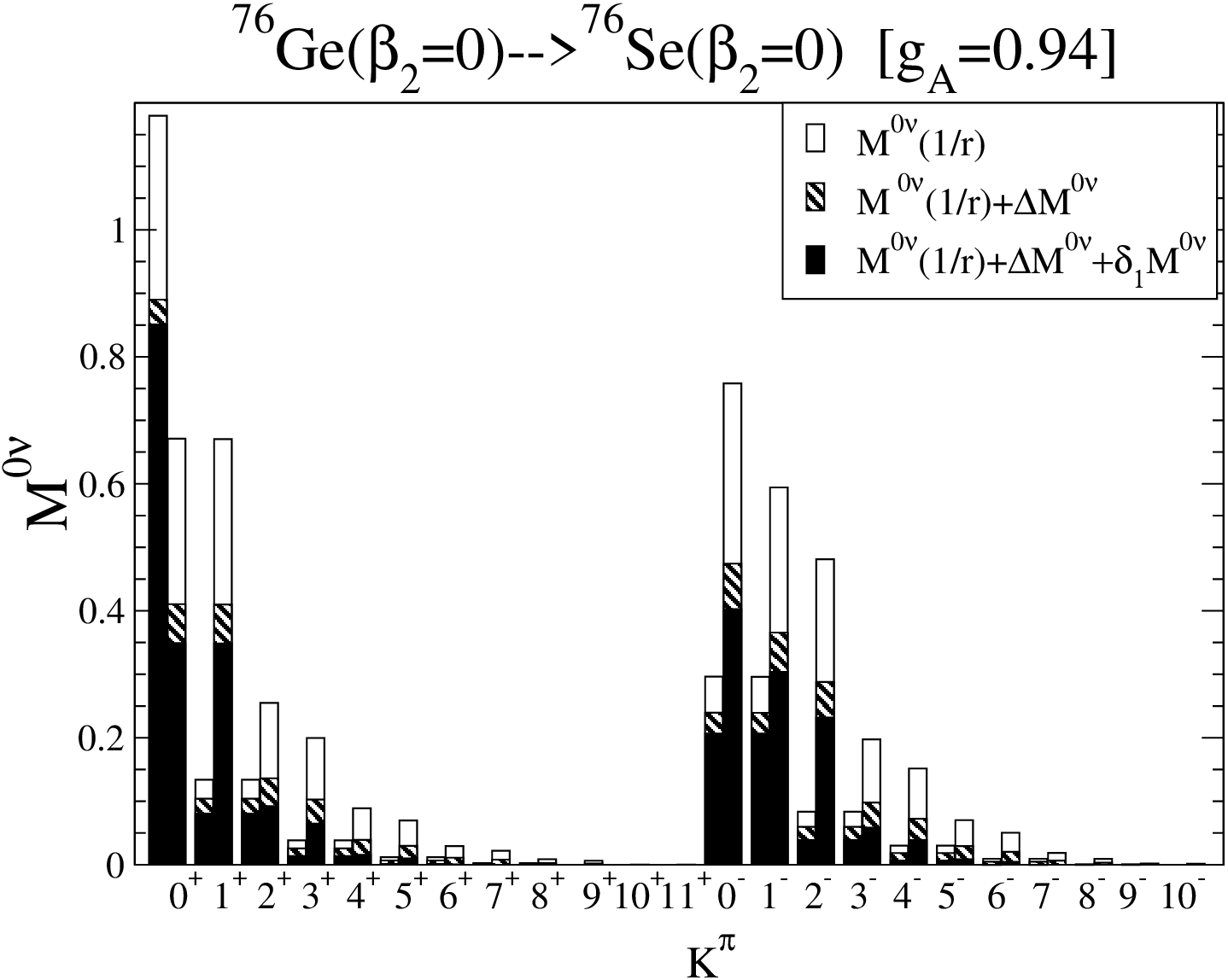}
\caption{ The same as in Fig.~\ref{fig.2} but for $^{76}$Ge$\rightarrow ^{76}$Se.}
\label{fig.3}
\end{figure}

The final results for the NME for $0\nu\beta\beta$ decays $^{76}$Ge$\rightarrow ^{76}$Se, $^{150}$Nd$\rightarrow ^{150}$Sm, $^{160}$Gd$\rightarrow ^{160}$Dy
are listed in Table~\ref{MEres} (now the values of the BCS overlap from the last column of
Table~\ref{table.1} are taken into account). In columns 4 and 9 the leading contributions $M_F(1/r)$ and $M_{GT}(1/r)$ are shown. In columns 5 and 10 the 
total corrections $\Delta M_F$ and $\Delta M_{GT}$ to $M_F(1/r)$ and $M_{GT}(1/r)$
are listed. 
One sees that these corrections reduce the results of the leading order by about
$20\%$ for the Fermi part and $40\%$ for the GT part, that agrees with the previous spherical QRPA calculations~\cite{Sim99}. 
In columns 6,7 and 11,12 the 
corrections $\delta_{1,2} M_F$ and $\delta_{1,2} M_{GT}$, respectively, coming from different choices of the s.r.c., are listed. In columns 8 and 13  both the F and GT parts of the total NME $M^{0\nu}_{\ell}=M^{0\nu}_{\ell}(1/r)+\Delta M^{0\nu}_{\ell}+\delta_2 M^{0\nu}_{\ell}$ are shown 
(we prefer here the final value of $M^{0\nu}$ corresponding to the modern self-consistent treatment of the s.r.c.~\cite{Sim09}).
Finally, in columns 14 and 15 the $0\nu\beta\beta$-decay NME $M'^{0\nu}$~(\ref{M'}) and corresponding decay half-lives (assuming $m_{\beta\beta}$=50 meV) are listed.
The corresponding $K^\pi$-decompositions of $M^{0\nu}$ are shown in Figs.~\ref{fig.2},\ref{fig.3}.
 
By inspecting Table~\ref{MEres} and Figs.~\ref{fig.2},\ref{fig.3}, one sees that the
difference between the spherical and deformed results mainly come from
the BSC overlap between the ground states of the initial and final nuclei. 
As for the $g_{pp}$ dependence of the $0\nu\beta\beta$-decay NME, 
it is much less pronounced as compared with the dependence of the amplitude of $2\nu\beta\beta$ decay.
A marked reduction of the total $M'^{0\nu}$ for the quenched value of $g_A$
can be traced back to a smaller prefactor $(g_A/1.25)^2$ in the definition of $M'^{0\nu}$~(\ref{M'}).

\begin{table}[h]
\centering
\caption{Different contributions to the total calculated NME $M'^{0\nu}$ for $0\nu\beta\beta$ decays $^{76}$Ge$\rightarrow ^{76}$Se, $^{150}$Nd$\rightarrow ^{150}$Sm, $^{160}$Gd$\rightarrow ^{160}$Dy. The BCS overlap is taken into account. In columns 4 and 9 the leading contributions $M_F(1/r)$ and $M_{GT}(1/r)$ are shown. In columns 5 and 10 the 
total corrections $\Delta M_F$ and $\Delta M_{GT}$ to $M_F(1/r)$ and $M_{GT}(1/r)$
are listed. In columns 6,7 and 11,12 the 
corrections $\delta_{i} M_F$ and $\delta_{i} M_{GT}$, respectively, coming from different choices of the s.r.c., are listed. In columns 8 and 13  both the F and GT parts of the total NME (\ref{M00}) are shown (we prefer here the final value of $M^{0\nu}$ corresponding to the modern self-consistent treatment of the s.r.c.~\cite{Sim09}).
Finally, in columns 14 and 15 the $0\nu\beta\beta$-decay NME $M'^{0\nu}$~(\ref{M'}) and corresponding decay half-lives (assuming $m_{\beta\beta}$=50 meV) are listed. }
\label{table.2}
\begin{tabular}{|c | c| c |c c c c c | c c c c c | c| c|}
	\hline
& & & & & & & & & & & & & & \\
$A$ & Def. &$g_A$ & \multicolumn{5}{c|}{$M^{0\nu}_F$} & \multicolumn{5}{c|}{$M^{0\nu}_{GT}$}&${M'}^{0\nu}$& $T^{0\nu}_{1/2}$, $10^{26}$ y
\\[2pt]
& & & $M(1/r)$&$\Delta M$& $\delta_1 M$& $\delta_2 M$& tot
& $M(1/r)$&$\Delta M$&$\delta_1 M$& $\delta_2 M$& tot
& & ($m_{\beta\beta}$=50 meV)
\\[4pt]	
\hline
76 & ``1" & 1.25 
& -2.83 & 0.69 & 0.40 &-0.08 & -2.22 
& 5.59 &-2.49&-0.88& 0.18 & 3.27 
& 4.69 
& 7.15\\
          &    & 0.94 
& -2.98 & 0.73 & 0.40 & -0.18 & -2.44 
& 7.69& -3.65& -1.47 & 0.27 & 4.31 
& 4.00 
& 9.83\\ 
 & ``0" & 1.25 
& -3.15 & 0.78 & 0.45 & -0.09& -2.47
& 6.37  &-2.85 & -1.01& 0.20 & 3.72
& 5.30 
& 5.60\\                      
   &           & 0.94 
&-3.31  & 0.82 & 0.46 & -0.10 & -2.59 
&7.28   &-3.15 & -1.04& 0.21 & 4.35 
& 4.10
& 9.36\\   
\hline
\hline
150 & ``1" & 1.25 
& -2.09 & 0.51 & 0.33 &-0.06 & -1.64 
& 4.01 &-1.86&-0.72& 0.14 & 2.29 
& 3.34 
& 0.41 \\
  &            & 0.94 
& -2.16 & 0.52 & 0.33 & -0.06 & -1.70 
& 4.44& -2.00& -0.73 & 0.14 & 2.58 
& 2.55 
& 0.71\\ 
 & ``0" & 1.25 
& -4.07 & 0.99 & 0.67 & -0.13& -3.21
& 7.35  &-3.54 & -1.46& 0.26 & 4.07
& 6.12 
& 0.12\\                      
       &       & 0.94 
&-4.12  & 1.00 & 0.68 & -0.13 & -3.25 
&7.69   &-3.65 & -1.47& 0.27 & 4.31 
& 4.52 
& 0.23\\   
\hline
\hline
160 & ``1"  & 1.25 
& -2.14 & 0.51 & 0.32 &-0.07 & -1.69 
& 4.57 &-2.04&-0.71& 0.14 & 2.67 
& 3.76 
& 2.26\\
\hline
\end{tabular}
\label{MEres}
\end{table}

The strongest effect of deformation on $M^{0\nu}$ (the suppression by about 40\% as compared to our previous QRPA result obtained with neglect of deformation) is found  in the case of $^{150}$Nd. This suppression can be traced back to a rather large difference in deformations of the ground states of $^{150}$Nd and $^{150}$Sm. Such an effect has been observed also by other authors~\cite{men09,Hir08,Bar09}. The modern self-consistent treatment of the s.r.c.~\cite{Sim09} makes the resulting $M'^{0\nu}=3.34$ (without quenching), even a bit larger than the NME $M'^{0\nu}=3.16$ of Ref.~\cite{Fang10} where the influence of the s.r.c. was completely neglected. This translates to the half-life $T^{0\nu}_{1/2}=4.1\cdot10^{25}$ y for the effective Majorana neutrino mass $\langle m_{\beta\beta} \rangle$ = 50 meV (cf. with $T^{0\nu}_{1/2}=4.60\cdot10^{25}$ y of Ref.~\cite{Fang10}). In the case of quenched $g_A$, the half-life $T^{0\nu}_{1/2}=7.1\cdot10^{25}$ y is about twice longer as a consequence of a smaller NME $M'^{0\nu}=2.55$.

\section{Conclusions}

In this paper a microscopic state-of-the-art approach 
to calculation of the nuclear matrix element $M^{0\nu}$ for neutrinoless double beta decay  with an account for nuclear deformation is presented in length and applied to calculate $M^{0\nu}$ for $^{76}$Ge, $^{150}$Nd and $^{160}$Gd.
The QRPA with a realistic residual interaction (the Brueckner $G$ matrix derived from the Bonn-CD nucleon-nucleon potential) is used as the underlying nuclear structure model.  The effects of the short range correlations and the quenching of the axial vector coupling constant $g_A$ are analyzed and found to be in accord with the spherical QRPA calculations.
The strongest effect of deformation on $M^{0\nu}$ (the suppression by about 40\% as compared to our previous QRPA result obtained with neglect of deformation) is found  in the case of $^{150}$Nd. This suppression can be traced back to a rather large difference in deformations of the ground states of $^{150}$Nd and $^{150}$Sm, that agrees with results by other authors.
The preliminary conclusion of Ref.~\cite{Fang10} that 
neutrinoless double beta decay of $^{150}$Nd provides one of the best probes of the Majorana neutrino mass is confirmed.

\acknowledgments
The authors acknowledge the support of the Deutsche Forschungsgemeinschaft under both SFB TR27 "Neutrinos and Beyond" and Graduiertenkolleg GRK683. The work of F.~\v{S}. was also partially supported by the VEGA Grant agency under the contract No.~1/0249/03.

\end{document}